\documentclass[final,5p,times,twocolumn]{elsarticle} 
\usepackage{hyperref}

\journal{Journal of \LaTeX\ Templates}

\begin{document}

\begin{frontmatter}

\title{Characterisation of a new RPC prototype using conventional gas mixture}
%\tnotetext[mytitlenote]{Fully documented templates are available in the elsarticle package on \href{http://www.ctan.org/tex-archive/macros/latex/contrib/elsarticle}{CTAN}.}

%% Group authors per affiliation:
\author[mymainaddress]{A. Sen\corref{mycorrespondingauthor}}
\cortext[mycorrespondingauthor]{Corresponding author}
\ead{arindam@jcbose.ac.in}
\address[mymainaddress]{Department of Physics, Bose Institute, Kolkata-700091, INDIA}
%\fntext[myfootnote]{Since 1880.}

%% or include affiliations in footnotes:
\author[mymainaddress]{S. Chatterjee}
%\author[mymainaddress]{R. Das}
\author[mymainaddress]{S. Das}
\author[mymainaddress]{S. Biswas}
%\ead[url]{www.elsevier.com}

%\author[mysecondaryaddress]{Global Customer Service\corref{mycorrespondingauthor}}
%\cortext[mycorrespondingauthor]{Corresponding author}
%\ead{support@elsevier.com}

%\address[mymainaddress]{1600 John F Kennedy Boulevard, Philadelphia}
%\address[mysecondaryaddress]{360 Park Avenue South, New York}

\begin{abstract}
Resistive Plate Chamber (RPC) is a well-known gaseous detector in the field of High Energy Physics (HEP) experiments for its good tracking capability, high efficiency, good time resolution, and low cost of fabrication. The main issue in RPC is its limitation in the rate handling capability. Several experimental groups have developed sophisticated techniques to increase the particle rate capability and reduce the noise rate of this detector. In this article, we discussed a new method for linseed oil coating in case of bakelite RPC detector to achieve good efficiency and the results obtained using a conventional gas mixture.
\end{abstract}


\end{frontmatter}

\section{Introduction}

HEP experiments have been using RPCs for triggering and tracking for their high efficiency and good time resolution. RPCs are also being used in several cosmic ray experiments to cover large detection area \cite{cosmic, cosmic2}. Future heavy-ion experiments $e.g.$ CBM will use RPCs for muon detection \cite{CBM}. RPC is a gaseous detector made up of resistive electrode plates $e.g.$ bakelite, glass, ceramic, etc \cite{rpc_2020, glass_rpc, ceramic_rpc}.

In bakelite RPC linseed oil coating \cite{BH06} is done to get rid of surface roughness issue \cite{CL09}. The linseed oil coating also helps to reduce the noise rate of the detector, protects the electrode plate from Hydrofluoric Acid (HF) corrosive effect and it also has photon quenching properties that reduce the UV sensitivity of the electrode plates.

In convensional linseed oil coated bakelite RPC a serious problem was observed in the BaBar experiment. The coated linseed oil formed stalagmite, that subsequently forms the conducting paths through the gas gap, around the spacers and the discharge permanently damages the detector. The formation of stalagmite is due to the polymerization of uncured linseed oil droplets present on the surface \cite{JV03}. However, lot of R\&D is performed and the solution is found out for this issue \cite{anuli}.

A new technique of linseed oil coating \cite{sen_2022} is introduced in the bakelite plates in which one can check visually whether there is any uncured oil present inside the gas gap. With this new technique one RPC prototype was built and tested with 100\% C$_{2}$H$_{2}$F$_{4}$ gas. In this article the test results of the RPC prototype with conventional C$_{2}$H$_{2}$F$_{4}$ and i-C$_{4}$H$_{10}$ mixture is presented.

\section{Detector description and experimental set-up}

The detector is built with two 27~cm$\times$~27~cm bakelite plates of thickness 2~mm and having bulk resistivity of $\sim$~3$\times$~10$^{10}$~$\Omega$~cm (at 22$^{\circ}$C). Four edge spacers, two gas nozzles, one button spacer made of polycarbonate (resistivity $\sim$~10$^{15}$~$\Omega$~cm) and having thickness of 2~mm are used to make the gas gap. The surface resistivity of the outer graphite surface of the electrode plates are measured to be $\sim$~510~k$\Omega$/$\Box$ and 540~k$\Omega$/$\Box$.

The signal is read out from the copper strips of dimension 2.5~cm$\times$~27~cm. The strips are covered with 100~$\mu$m thick mylar foils to isolate them from the graphite layers. The details of the fabrication of the chamber is described in \cite{sen_2022}.

\begin{figure}[htb!]
	\centering{
		\includegraphics[scale=0.32]{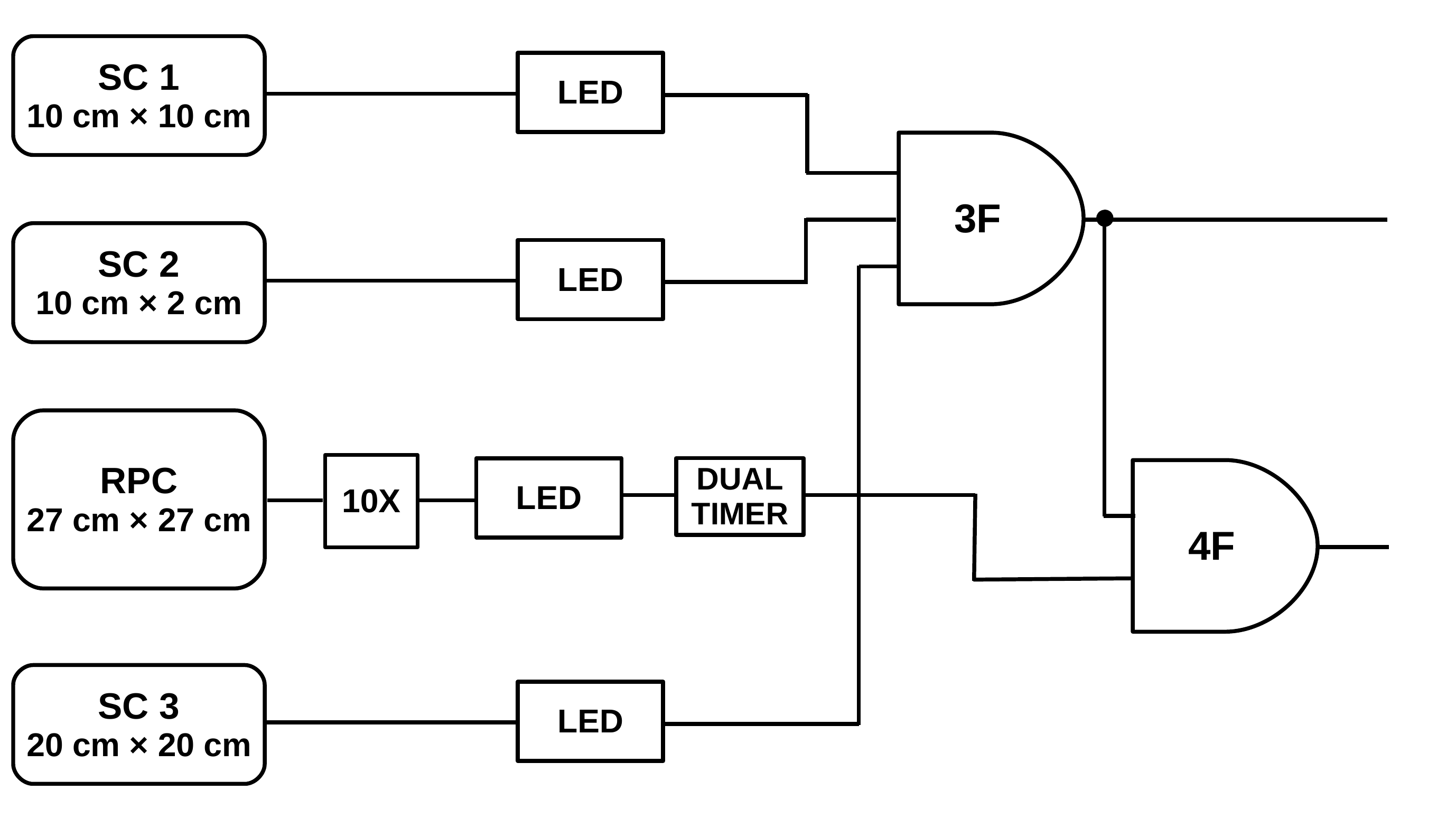}
	}
	\caption{Block diagram of the cosmic ray test set-up for characterization of the detector \cite{sen_2022}.}\label{ckt}
\end{figure}

Three scintillation detectors of dimensions 10~cm~$\times$~10~cm (SC1), 10~cm~$\times$~2~cm (SC2), 20~cm~$\times$~20~cm (SC3) are used to generate the trigger for the detector. All the scintillators are operated at +1550~V and -15~mV threshold is applied to the leading edge discriminator (LED). The RPC signal from the pick-up strip is first fed to a 10x fast amplifier and then the output goes to the LED. Suitable thresholds are applied to the LEDs to reduce the noise. From the LED, one output goes to the scalar to count the number of the signal from the RPC which is known as the noise count or singles count of the chamber. The other output from LED goes to the dual timer where the discriminated RPC signal is stretched to avoid any double counting of the pulses and also to apply the proper delay to match the signal with the trigger. The output of the dual timer is taken in coincidence with the trigger and this is defined as the 4-fold signal. The window of the cosmic ray test set-up is of area 10~cm~$\times$~2~cm. The block diagram of the set-up is shown in Figure~\ref{ckt}.

%%%%%%%%%%%%%%%%%%%%%%%%%%%%%%%%%%%%%%%%%%%%%%%%%	

%%%%%%%%%%%%%%%%%%%%%%%%%%%%%%%%%%%%%%%%%%%%%%%%%

\section{Result}

\begin{figure}[htb!]
	\centering{
		\includegraphics[scale=0.46]{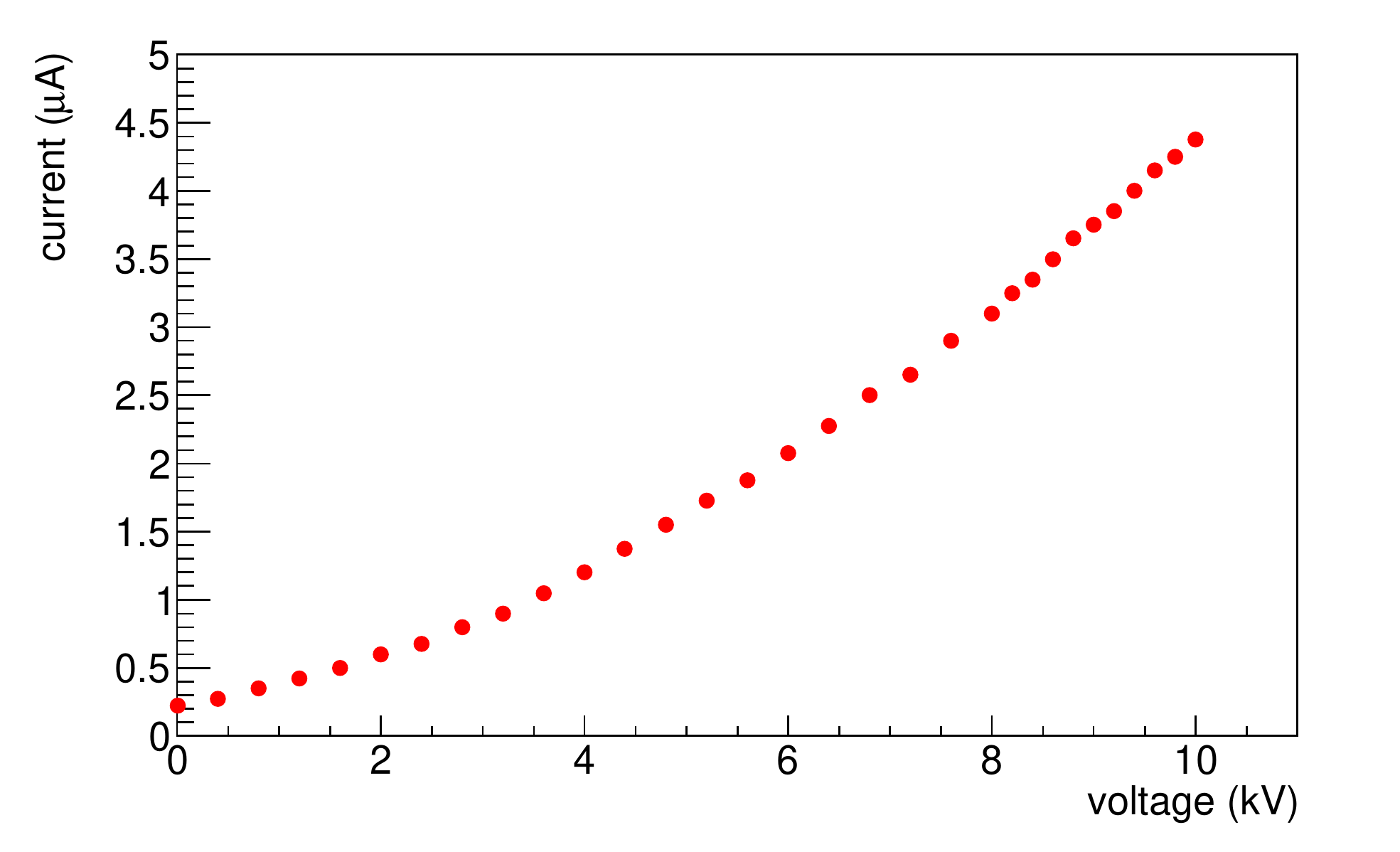}
	}
	\caption{Leakage current as a function of the voltage for C$_{2}$H$_{2}$F$_{4}$ and i-C$_{4}$H$_{10}$ gas mixture in 90/10 volume ratio (colour online).}\label{iv}
\end{figure}

The chamber is tested with cosmic rays in avalanche mode. The detector is purged with Tetrafluoroethane (C$_{2}$H$_{2}$F$_{4}$) and Isobutane (i-C$_{4}$H$_{10}$) gas mixture in 90/10 volume ratio. The leakage current through the RPC module is measured as a function of the applied high voltage (HV) and shown in Figure~\ref{iv}.

%%%%%%%%%%%%%%%%%%%%%%%%%%%%%%%%%%%%%%%%%%%%%%%%%	

%%%%%%%%%%%%%%%%%%%%%%%%%%%%%%%%%%%%%%%%%%%%%%%%%

The efficiency of the RPC module for the cosmic rays is defined as the ratio of the 4-fold counts to the 3-fold coincidence trigger count of the plastic scintillator telescope for a fixed duration and the noise rate of the RPC, is defined as the number of counts per unit area of the strip per second. The efficiency and noise rate are studied by varying the applied HV. The RPC is tested with -~20~mV and -~25~mV threshold settings to the LED. An efficiency of greater than 90\% is achieved from 10~kV onwards for both the threshold settings. The maximum noise rates are found to be 120 Hz/cm$^{2}$ and 80 Hz/cm$^{2}$ for the -~20~mV and -~25~mV thresholds respectively.

%%%%%%%%%%%%%%%%%%%%%%%%%%%%%%%%%%%%%%%%%%%%%%%%%	

\begin{figure}[htb!]
	\centering{
		\includegraphics[scale=0.495]{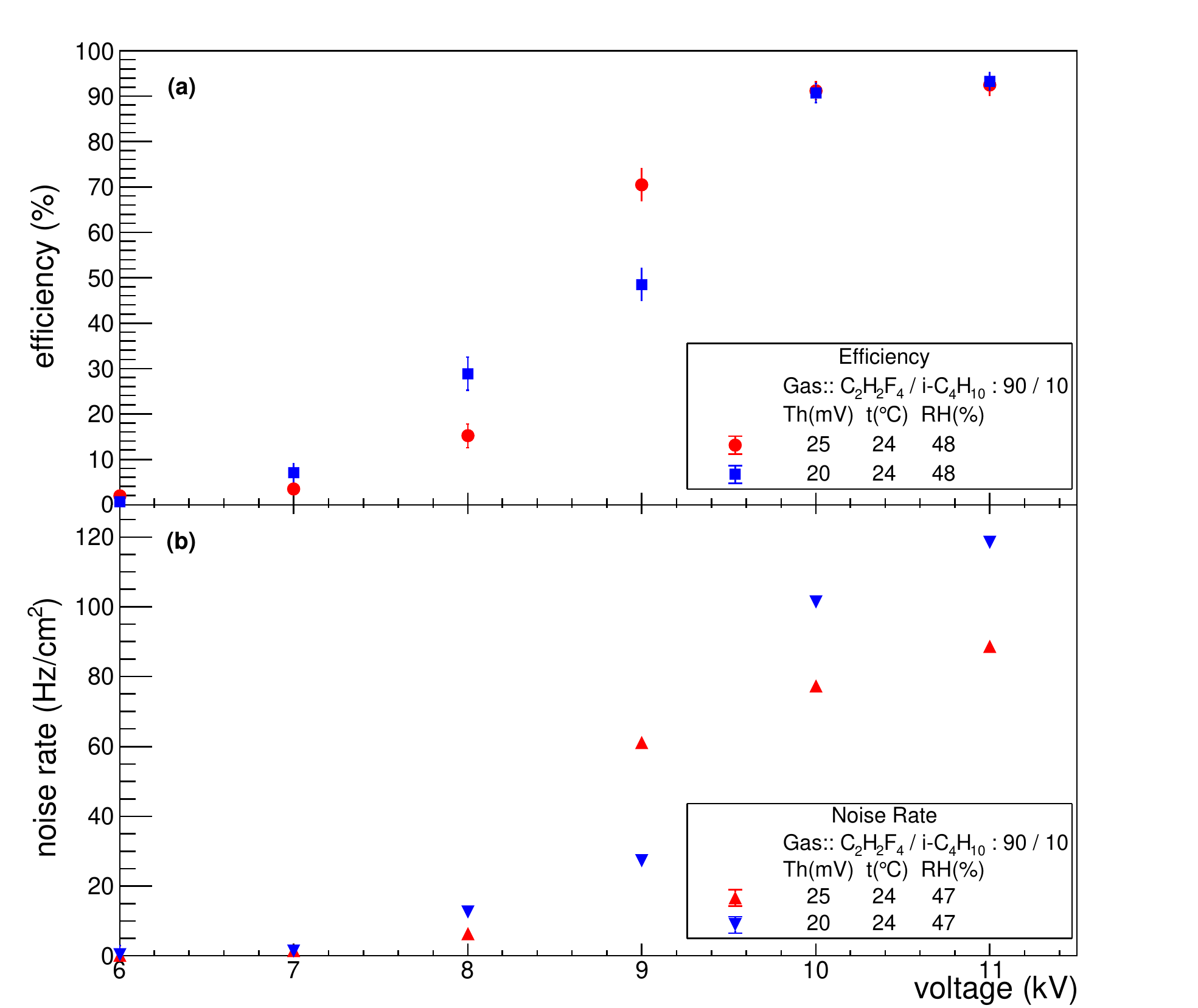}
	}
	\caption{(a) Efficiency as a function of the applied voltage, (b) Noise rate as a function of the applied voltage for a gas mixture of C$_{2}$H$_{2}$F$_{4}$ and i-C$_{4}$H$_{10}$ in the 90/10 volume ratio (colour online).}\label{eff_noise_oil}
\end{figure}
%%%%%%%%%%%%%%%%%%%%%%%%%%%%%%%%%%%%%%%%%%%%%%%%%

\section{Summary}

A linseed oil-coated RPC prototype of dimension 27~cm~$\times$~27~cm is fabricated using indigenous resistive bakelite plates. The linseed oil coating is done before making the gas gap. Before building the detector, electrode plates are checked visually whether the oil is cured properly or if any uncured oil is present on the surface.

The detector is tested with Tetrafluoroethane (C$_{2}$H$_{2}$F$_{4}$) and Isobutane (i-C$_{4}$H$_{10}$) in 90/10 volume ratio. Both the current and noise rate are very low for this gas mixture compared to the 100\% C$_{2}$H$_{2}$F$_{4}$ used earlier for the same detector \cite{sen_2022}. An efficiency of greater than 90\% is found from 10~kV onwards with a maximum noise rate of 120~Hz/cm$^{2}$ at -~20~mV threshold.

\section{Acknowlegement}

The authors would like to thank Ms. Rudrapriya Das for her help in running the experiment and Prof. Sanjay K Ghosh for many valuable discussions and suggestions during the course of study. The authors would also like to thank Mr. Subrata Das for building the readout strips. This work is partially supported by the CBM-MuCh project from BI-IFCC, DST, Govt. of India. A. Sen acknowledges his Inspire Fellowship research grant [DST/INSPIRE Fellowship/2018/IF180361].

%\bibliography{mybibfile}

\end{document}